\documentclass{statsoc}

\usepackage[a4paper]{geometry}
\usepackage{graphicx}
\graphicspath{ {image/} }
\usepackage{amsmath}
\usepackage{natbib}
\usepackage{bm}
\usepackage{multirow}
\usepackage{enumerate,multicol}
\usepackage{booktabs}
\usepackage{amsfonts}
\usepackage{rotating}
\usepackage{hyperref}
\usepackage[titletoc,title]{appendix}
\usepackage{etoolbox}
\usepackage{float}
\usepackage[T1]{fontenc}
\usepackage[utf8]{inputenc}
\usepackage{afterpage}
\usepackage[caption=false]{subfig}
\usepackage{amssymb}
\usepackage{mathtools}
\usepackage{makecell}\usepackage{makecell}
\usepackage[titletoc,title]{appendix}
\usepackage{fancyvrb} 
\usepackage{listings,color} 
\usepackage{moreverb}

\makeatletter
\patchcmd{\@makecaption}
  {\parbox}
  {\advance\@tempdima-\fontdimen2} 
  {}{}
\makeatother 

\makeatletter
\newcommand{\vast}{\bBigg@{4}}
\newcommand{\Vast}{\bBigg@{5}}
\makeatother

\title[A Longitudinal Model to Handle Missingness in Economic Evaluations]{A Bayesian Parametric Approach to Handle Missing Longitudinal Outcome Data in Trial-Based Health Economic~Evaluations}
\author[A. Gabrio]{Andrea Gabrio}
\address{Department of Statistical Science, University College London
London,
UK.}
\email{ucakgab@ucl.ac.uk}
\author[MJ. Daniels]{Michael J. Daniels}
\address{Department of Statistics, University of Florida,
Gainesville,
FL.}
\author[A. Gabrio, MJ. Daniels and G. Baio]{Gianluca Baio}
\address{Department of Statistical Science, University College London
London,
UK.}

\newcommand\ack{\section*{Acknowledgements}}

\begin{document}
\begin{abstract}
Trial-based economic evaluations are typically performed on cross-sectional variables, derived from the responses for only the completers in the study, using methods that ignore the complexities of utility and cost data (e.g.~skewness and spikes). We present an alternative and more efficient Bayesian parametric approach to handle missing longitudinal outcomes in economic evaluations, while accounting for the complexities of the data.
We specify a flexible parametric model for the observed data and partially identify the distribution of the missing data with partial identifying restrictions and sensitivity parameters. We explore alternative nonignorable scenarios through different priors for the sensitivity parameters, calibrated on the observed~data. 
Our approach is motivated by, and applied to, data from a trial assessing the cost-effectiveness of a new treatment for intellectual disability and challenging behaviour.
\end{abstract}

\keywords{Bayesian Statistics; Cost-Effectiveness; Longitudinal Data; Missing Data; Sensitivity Analysis.}

\section{Introduction}\label{intro}
Economic evaluation alongside Randomised Clinical Trials (RCTs) is an important and increasingly popular component of the process of technology appraisal \citep{NICE2013}. The typical analysis of individual level data involves the comparison of two interventions for which suitable measures of clinical benefits and costs are observed on each patient enrolled in the trial at different time points throughout the follow~up. 

Typically, clinical benefits are measured through multi-attribute utility instruments (e.g.~the EQ-5D-3L: \url{http://www.euroqol.org}), costs are obtained from clinic resource records and both are summarised into cross-sectional quantities, e.g.~Quality Adjusted Life Years (QALYs).  The main objective of the economic analysis is a) to combine the population average clinical benefits (or effectiveness) and costs in order to determine the most ``cost-effective'' intervention, given current evidence; and b) to assess the impact of the uncertainty in the model inputs on the decision-making process
\citep{Claxtonb,Briggs2000,Spiegelhalterb,OHagan2004,Sculpher,Briggs2006,Jackson,Baioa}.

Individual level data from RCTs are almost invariably affected by missingness. The recorded outcome process is often incomplete due to individuals who drop out or are observed intermittently throughout the study, causing some observations to be missing. In most applications, the economic evaluation is performed on the cross-sectional variables, computed using only the data from the individuals who are observed at each time point in the trial (completers), with at most limited sensitivity analysis to missingness assumptions~\citep{Noble2012,Gabrio,Leurent2018}. This, however, is an extremely inefficient approach as the information from the responses of all partially observed subjects is completely lost and it is also likely biased unless the completers are a random sample of the subjects on each arm~\citep{Little2002}. 

Handling missingness can be very challenging, especially because missing observations may themselves provide information about the distribution of the unobserved data \citep{Rubina,Little2002}. Dealing with informative missingness is not straightforward as inference can be drawn only under untestable assumptions about the unobserved data and is often sensitive to the particular assumptions made \citep{Molenberghs1997}. It is therefore desirable to assess the robustness of the inference by varying these assumptions in a principled way \citep{Scharfstein1999,Vansteelandt2006,Daniels2008}.

The problem of informative missingness is often embedded within a more complex framework, which makes the modelling task in economic evaluations particularly challenging. Specifically, the effectiveness and cost data typically present a series of complexities that need to be simultaneously addressed to avoid biased results. First, the presence of a bivariate outcome requires the use of appropriate methods that deal with correlation \citep{OHagan}. Second, outcome data typically have empirical distributions that are highly skewed. The adoption of  parametric distributions that can account for skewness (e.g.~Beta for the utilities and Gamma or Log-Normal for the costs) has been suggested to improve the fit \citep{Nixon,Thompson}. In addition, data may exhibit spikes at one or both of the boundaries of the range for the underlying distribution that may induce high skewness in the data that is difficult to capture using standard parametric models~\citep{Cooper}. For example, some patients in a trial may not accrue any cost at all or some individuals may be associated with perfect health, i.e.~unit QALY. The use of more flexible formulations, known as \textit{hurdle models}, explicitly accounts for these ``structural'' values.  Hurdle models are essentially a mixture between a point mass distribution (the spike) and a parametric model fit to the natural range of the relevant variable without the boundary values. Hurdle models have been applied in economic evaluations for handling either costs or QALYs~\citep{Baio2014,Gabriob}.

Using a recent randomised trial as our motivating example, we present a Bayesian parametric model for conducting inference on a bivariate health economic longitudinal response. We specify our model to account for the different types of complexities affecting the data while accommodating a sensitivity analysis to explore the impact of alternative missingness assumptions on the inferences and on the decision-making process for health technology assessment.

\subsection{Positive Behaviour Support Trial}\label{PBS}
The Positive Behaviour Support (PBS) study~\citep{Hassiotis2018} is a multicenter randomised controlled trial that, among its objectives, aimed to evaluate the cost-effectiveness of a new multicomponent intervention (PBS, 108 subjects) relative to treatment as usual (TAU, 136 subjects) for individuals suffering from mild to severe intellectual disability and challenging behaviour. The primary instruments used to assess the clinical benefits and costs were the EQ-5D-3L questionnaires and family/paid carer clinic records, respectively. Utilities are derived from the health questionnaires using the time trade-off algorithm \citep{NICE2013} and are defined on the interval $[-0.594,1]$, where 1 represents the perfect health state while negative values indicate states that are considered ``worse than death''. Costs, expressed in~$\pounds{}$, are obtained from the clinic records. Subjects associated with either or both a utility of one and a null cost are observed throughout the study. Measurements were scheduled to be collected at baseline and at 6 and 12 months after baseline.

Let $\bm u_i=(u_{i0},\ldots,u_{iJ})$ and $\bm c_i=(c_{i0},\ldots,c_{iJ})$ denote the vectors of utilities and costs that were supposed to be observed for subject $i$ at time $j$ in the study, with~$j~\in~\{0,1,J=2\}$. We denote with $\bm y_{ij}=(u_{ij},c_{ij})$ the bivariate outcome for subject $i$ formed by the utility and cost pair at time $j$. Both outcomes were partially observed and missingness was nonmonotone in the sense that if $\bm y_{ij}$ was unobserved then $\bm y_{i\hspace*{0.05mm}j+1}$ could be either observed or unobserved. We group the individuals according to the missingness patterns and denote with $\bm r_{ij}=(r^u_{ij},r^c_{ij})$ a pair of indicator variables that take value $1$ if the corresponding outcome for subject $i$ at time $j$ is observed and $0$ otherwise. We denote with $\bm r_i=(\bm r_{i0};\bm r_{i1};\bm r_{i2})$ the missingness pattern to which subject $i$ belongs, where each pattern is associated with different  values for $\bm r_{ij}$. For example, the pattern $\bm r=\bm 1$ is associated with the set $\bm r=(1,1;1,1;1,1)$ and corresponds to the completers pattern. We denote with~$R_t$ the total number of observed patterns either in the control ($R_1=5$) or intervention ($R_2=9$) group.
Table~\ref{patterns} reports the missingness patterns in each treatment group as well as the number of individuals and the observed mean responses within each~pattern. 

\begin{center}
TABLE 1 HERE
\end{center}

The number of observed patterns is relatively small and with the exception of the completers ($\bm r= \bm 1$) the patterns are quite sparse. Baseline costs in both treatment groups are the only fully observed variables, while the average proportion of missing utilities and costs is respectively $21\%$ and $10\%$ for the control $(t=1)$ and $10\%$ and $8\%$ for the~intervention~$(t=2)$. 

\subsection{Standard Approach to Economic Evaluation}\label{standard}
To perform the economic evaluation, aggregated measures for both utilities and costs are typically derived from the longitudinal responses recorded in the study. QALYs ($e_{it}$) and total costs ($c_{it}$) measures are computed as:
\begin{equation}\label{QALYs}
e_{it}=\sum_{j=1}^{J}(u_{ijt}+u_{i\! j-1\hspace{0.005em}t})\frac{\delta_{j}}{2} \;\;\; \text{and} \;\;\;\  c_{it}=\sum_{j=1}^{J}c_{ijt},
\end{equation} 
where $\delta_{j}=\frac{\text{Time}_{j}-\text{Time}_{j-1}}{\text{Unit of time}}$ is the fraction of the time unit (12 months, in the PBS study) between consecutive measurements. The economic evaluation is then carried out by applying some parametric model $p(e_{it},c_{it}\mid \bm \theta)$ to these cross-sectional quantities, typically using linear regression methods to account for the imbalance in some baseline variables between treatments \citep{Manca,Asselt,Agency}. Finally, QALYs and total costs population mean~values are derived from the model:
\begin{equation}\label{mu}
\mu_{et} = \text{E}\left(e_{it} \mid \bm \theta\right) \;\;\; \text{and} \;\;\; \mu_{ct} = \text{E}\left(c_{it} \mid \bm \theta \right). 
\end{equation} 
The quantities $\mu_{et}$ and $\mu_{ct}$ represent the target of interest in each treatment group $t$ and are used in assessing the relative cost-effectiveness of the interventions.

In the original economic evaluation of the PBS study, the quantities in Equation~\ref{QALYs} were derived based on the longitudinal responses for only the completers, while discarding all other partially observed data. Next, the quantities in Equation~\ref{mu} were obtained under a frequentist approach in which the two outcome variables $(e_{it},c_{it})$ were modelled independently assuming normality for the underlying distributions and using linear regression methods to control for differences in baseline~values.

The modelling approach used in the original analysis has the limitation that $\mu_{et}$ and $\mu_{ct}$ are derived based only on the completers in the study and does not assess the robustness of the results to a range of plausible missingness assumptions. The model also fails to account for the different complexities that affect the utility and cost data in the trial: from the correlation between variables to the skewness and the presence of structural values (zero for the costs and one for the utilities) in both~outcomes.

\subsection{A Longitudinal Model to Deal with Missingness}\label{longitudinal}
We propose an alternative, more efficient and less biased approach to deal with a missing bivariate outcome in economic evaluations while simultaneously allowing for the different complexities that typically affect utility and cost data. Our approach includes a longitudinal model that improves the current practice by taking into account the information from all observed data as well as the time dependence between the responses. 
The targeted quantities can then be obtained by applying the same formulae in Equation~\ref{QALYs} to the marginal means at each time for $\bm y_{ij}=(u_{ij},c_{ij})$, which can be easily derived from the model. This can be accomplished through the specification of a joint distribution $p(\bm y, \bm r \mid \bm \omega)$ for the response and missingness pattern, where $\bm \omega$ is some relevant parameter~vector.

We define the data as $\bm y=(\bm y_{obs},\bm y_{mis})$ to indicate the subsets that are observed and missing. Next, define $p(\bm y \mid \bm \theta)$ as the response model, parameterised by $\bm \theta$, and $p(\bm r \mid \bm y, \bm \psi)$ as the missingness model, with parameters $\bm \psi$. Missingness is said to be \textit{ignorable} if the following three conditions hold~\citep{Little2002}: (1) $p(\bm r \mid \bm y,\bm \psi)=p(\bm r \mid \bm y_{obs},\bm \psi)$, that is, missingness depends only on the observed responses, a condition known as Missing At Random (MAR); (2) the parameter $\bm \omega$ of the joint model $p(\bm y, \bm r \mid \bm \omega)$ can be decomposed as $(\bm \theta,\bm \psi)$, with $p(\bm y \mid \bm \theta)$ and $p(\bm r\mid \bm y,\bm \psi)$; (3) the parameters of the response and missingness model are a priori independent, that is $p(\bm \omega)=p(\bm \theta)p(\bm \psi)$.

When any of these conditions is not satisfied, missingness is said to be \textit{nonignorable}. Often, this is due to the failure of the first condition, which implies $p(\bm r \mid \bm y_{obs},\bm y_{mis}, \bm \psi)\neq p(\bm r \mid \bm y_{obs}, \bm y^\prime_{mis} \bm \psi)$ for $\bm y_{mis} \neq \bm y^\prime_{mis}$, known as Missing Not At Random (MNAR). In this case, the joint model $p(\bm y,\bm r)$ will require untestable assumptions about the missing data in order to be identified. We specify our nonignorable modelling strategy using the extrapolation factorisation and a pattern-mixture approach with identifying restrictions \citep{Little1994,Linero2017}. 

In this work we present a parametric model with a fully Bayesian framework that can account for both skewness and structural values within a partially-observed outcomes setting. A major advantage of adopting a Bayesian approach is the ability to allow for the formal incorporation of external evidence into the analysis through the use of informative prior distributions. This is a crucial element for conducting sensitivity analysis to assess the robustness of the results to a range of plausible missing data~assumptions. 

\subsection{Outline}\label{outline}
In Section~\ref{model} we describe the general strategy used to define the model and the factorisation chosen to specify the joint distribution of the cost and utility data and the missingness patterns. In Section~\ref{wm} we introduce the parametric model implemented for the distribution of the observed data and present alternative specifications to identify the joint model under nonignorability. In Section~\ref{SA} we introduce the identifying restrictions used and the approach followed to conduct sensitivity analysis. In Section~\ref{results} we implement our model to draw inferences on the PBS study under alternative missingness assumptions. In Section~\ref{evaluation} we summarise the results under each scenario from a decision-maker perspective and compare the implications in terms of cost-effectiveness. We close in Section~\ref{conclusions} with a discussion.

\section{Modelling Framework}\label{model}
We define our modelling strategy following \citet{Linero2015} and factor the joint distribution for the response and missingness as:
\begin{equation*}
p(\bm y, \bm r \mid \bm \omega)=p(\bm y^{\bm r}_{obs}, \bm r \mid  \bm \omega)p(\bm y^{\bm r}_{mis} \mid \bm y^{\bm r}_{obs}, \bm r, \bm \omega)
\end{equation*}
where $\bm y^{\bm r}_{obs}$ and $\bm y^{\bm r}_{mis}$ indicate the observed and missing responses within pattern $\bm r$, respectively. This is the extrapolation \textit{factorisation} and factors the joint into two components, of which the extrapolation \textit{distribution} $p(\bm y^{\bm r}_{mis} \mid \bm y^{\bm r}_{obs}, \bm r, \bm \omega)$ remains unidentified by the data in the absence of unverifiable assumptions about the full data \citep{Daniels2008}. 
To specify the \textit{observed data distribution} $p(\bm y^{\bm r}_{obs}, \bm r \mid  \bm \omega)$ we use a \textit{working model} $p^{\star}$ for the joint distribution of the response and missingness \citep{Linero2015}.
\begin{equation*}
p(\bm y^{\bm r}_{obs}, \bm r \mid \bm \omega)=\int p^{\star}(\bm y, \bm{r} \mid  \bm \omega)d \bm y_{mis}
\end{equation*}
Since we use $p^{\star}(\bm y, \bm r \mid  \bm \omega)$ only to obtain a model for $p(\bm y^{\bm r}_{obs}, \bm r \mid  \bm \omega)$ and not as a basis for inference, the extrapolation distribution is left unidentified. Any inference depending on the observed data distribution may be obtained using the working model as the true model, with the advantage that  it is often easier to specify a model for the the full data $p(\bm y,\bm r)$ compared with a model  for the observed data~$p(\bm y^{\bm r}_{obs},\bm r)$.


We specify $p^*$ using a pattern mixture approach, factoring the joint $p(\bm y,\bm r \mid \bm \omega)$ as the product between the marginal distribution of the missingness patterns $p(\bm r\mid \bm \psi)$ and the distribution of the response conditional on the patterns $p(\bm y\mid \bm r,\bm \theta)$, respectively indexed by the distinct parameter vectors $\bm \psi$ and $\bm \theta$. If missingness is monotone it is possible to summarise the patterns by dropout time and directly model the dropout process \citep{Daniels2008,Gaskins}. Unfortunately, as it often occurs in trial-based health economic data, missingness in the PBS study is mostly nonmonotone and the sparsity of the data in most patterns makes it infeasible to fit the response model within each pattern, with the exception of the completers ($\bm r= \bm 1$). Thus, we decided to collapse together all the non-completers patterns ($\bm r \neq \bm 1$) and fit the model separately to this aggregated pattern and to the completers. The model can be represented as:
\begin{align*}\label{workm}
\begin{split}
    p(\bm y,\bm r \mid \bm \omega)={}& p(\bm r \mid \bm \psi ) \left[ p(\bm y \mid \bm r=\bm 1,\bm \lambda) \right]^{\mathbb{I}\{\bm r= \bm 1\}} \\
         & \left[ \prod_{\bm r\geq 2}p(\bm y^{\bm r}_{obs} \mid  \bm r ,\bm \eta) \right]^{ \mathbb{I}\{\bm r\neq \bm 1\} } 
\end{split}  \;\;\;\; \vast \} \;\;\;\;\;  \text{observed data distribution} \\
    {}& \left[ \prod_{\bm r\geq 2} p\left(\bm y^{\bm r}_{mis} \mid \bm y^{\bm r}_{obs},  \bm r,\bm \xi \right) \right]^{ \mathbb{I}\{\bm r\neq \bm 1\}}
    \;\;\;\;\;\;\;\;\;\; \text{extrapolation distribution}
\end{align*}
where $\bm \omega=(\bm \theta,\bm \psi)$, $\bm \lambda$ and $\bm \eta$ are the distinct subsets of $\bm \theta$ that index the response model in the completers and non-completers patterns, and $\bm \xi$ is the subset of~$\bm \eta$ that indexes the extrapolation distribution.
The joint distribution has three components. The first is given by the model for the patterns and the model for the completers ($\bm r= \bm 1$), where no missingness occurs. The second component is a model for the observed data in the collapsed patterns $\bm r \neq \bm 1$ that, together with the first component, form the observed data distribution. The last component is the extrapolation~distribution.

Because the targeted quantities of interest (Equation~\ref{mu}) can be derived based on the marginal utility and cost means at each time $j$, in our analysis we do not require the full identification of $p(\bm y^{\bm r}_{mis} \mid \bm y^{\bm r}_{obs},  \bm r,\bm \xi)$. Instead, we only partially identify the extrapolation distribution using \textit{partial identifying restrictions} \citep{Linero2017}. Specifically, we only require the identification of the marginal means for the missing responses in each pattern. 

Let $\mathcal{I}^{\bm r}$ be the indices of the missing observations in pattern $\bm r$ and let $\mathcal{J}_{\bm r}^{\bm{r}^\prime} \subseteq \mathcal{I}^{\bm r}$ be the subset of the indices in $\mathcal{I}^{\bm r}$ for which there are observed responses in~$\bm r^\prime$. We denote with $\bm y^{\bm r}_{mis}=\bm y^{\bm r}({\mathcal{I}^{\bm r}})$, the missing responses in pattern $\bm r$. Next, we denote with $\bm y_{obs}^{{\bm r}^\prime}({\mathcal{J}_{\bm r}^{{\bm r}^\prime}}) \subseteq \bm y^{\bm r^\prime}_{obs}$ the subset of the observed responses in ${\bm r}^\prime$ that corresponds to~$\bm y^{\bm r}_{mis}$.

We identify the marginal mean of $\bm y^{\bm r}_{mis}$ using the observed values $\bm y_{obs}^{\bm r^\prime}({\mathcal{J}_{\bm r}^{{\bm r}^\prime}})$, averaged across $\bm r^\prime \neq \bm1$, and some~\textit{sensitivity~parameters}~$\bm \Delta=(\Delta^u,\Delta^c)$. Therefore, we compute the marginal means by averaging only across the observed components in pattern ${\bm r}^\prime$ and ignore the components that are missing. 
\begin{equation*}
\mbox{E}\left[\bm{y}^{\bm r}_{mis} \mid \bm r \right]=\mbox{E}\left[ \underset{{\bm r}^\prime \neq \bm 1, \mathcal{J}^{{\bm r}^\prime}_{\bm r}}{\mbox{E}}\left[ \bm y_{obs}^{{\bm r}^\prime}({\mathcal{J}_{\bm r}^{{\bm r}^\prime}})+\bm \Delta \mid {\bm r}^\prime \right] \right].
\end{equation*}
Alternative identifying restrictions for nonmonotone missing data are reviewed in \citet{Linero2017}. 
We start by setting a benchmark assumption with~$\bm \Delta=\bm 0$, and then explore the sensitivity of the results to alternative scenarios by using different prior distributions on~$\bm \Delta$, calibrated on the observed data. 
Once the working model has been fitted to the observed data and the extrapolation distribution has been identified, the overall marginal mean for the response model can be computed by marginalising over $\bm r$, i.e.~$\text{E}\left[\bm Y\right]=\sum_{\bm r} p(\bm r)\text{E}\left[\bm Y \mid \bm r \right]$.
 
\section{Model for the missingness patterns and observed response}\label{wm}

The distribution of the number of patterns is a multinomial on $\{1,\ldots,R_t\}$, with the total number of patterns $R_t$ and the probabilities $\bm \psi^{\bm r}_t$ conditional on the treatment assignment~$t$. We specify a prior for $\bm \psi^{\bm r}_t$ that gives more weight on the completers pattern and equal weights to the other~patterns. Specifically, we choose a~$\text{Dirichlet}(1-x,\frac{x}{R^\star},\ldots,\frac{x}{R^\star})$~prior, where $x$ is the expected total dropout rate and $R^\star=64$ is the total number of potential patterns in the study. This is consistent with the design of the study, where the experimenter expects at least $(1-x)\%$ of the individuals to provide complete data, i.e.~to fall in $\bm r= \bm 1$. In practice, this prior is not likely to affect the results as the amount of observed data is enough to learn the posterior of $\bm \psi^{\bm r}_t$. For comparison purposes, we also consider another specification based on a noninformative $\text{Dirichlet}(1,\ldots,1)$~prior for $\bm \psi^{\bm r}_t$.  Posterior results are robust to the alternative prior choices.

The distribution of the observed responses $\bm y_{ijt}=(u_{ijt},c_{ijt})$ is specified in terms of a series of conditional distributions that capture the dependence between utilities and costs as well as the time dependence. We now drop the treatment indicator $t$ for clarity.
To account for the skewness we use Beta and Log-Normal distributions for the utilities and costs, respectively. Since the Beta distribution does not allow for negative values, we scaled the utilities on $[0,1]$ through the transformation $u^{\star}_{ij}=\frac{u_{ij}-\text{min}(\bm u_{j})}{\text{max}(\bm u_{j})-\text{min}(\bm u_{j})}$, and fit the model to these transformed variables. To ease the notation we refer to these quantities simply as~$u_{ij}$. 

To account for the structural values $u_{ij}=1$ and $c_{ij}=0$ we use a hurdle approach by including in the model the indicator variables $d^u_{ij}:=\mathbb{I}(u_{ij}=1)$ and $d^c_{ij}:=\mathbb{I}(c_{ij}=0)$, which take value $1$ if subject $i$ is associated with a structural value at time $j$ and 0 otherwise. The probabilities of observing these values, as well as the mean of each variable, are then modelled conditionally on the utilities and costs at the current and previous times via linear regressions defined on the logit or log~scale.  The model can be summarised as follows (for simplicity we omit the subject index~$i$). 

At time $j=0$, we model the nonzero costs $c_0\neq 0$ and the indicator~$d^c_0:=\mathbb{I}(c_0=0)$~as:
\begin{align*}
c_0\mid d^c_0=0 & \sim \text{LogNormal}\left(\nu_0^c,\tau^c_0\right)\\
d^c_0&\sim\mbox{Bernoulli}(\pi_0^c)
\end{align*}
where $\nu^c_0$ and $\tau^c_0$ are the mean and standard deviation for $c_0$ given $c_0\neq 0$ on the log scale, while $\pi^c_0$ is the probability of a zero cost value. 
We next model the utilities and the indicator $d^u_0:=\mathbb{I}(u_0=1)$ conditionally on the costs at the same~time:
\begin{align*}
u_0 \mid d^u_0=0, c_0& \sim \text{Beta}\left(\nu_0^u,\sigma^u_0\right)\\
\text{logit}(\nu^u_0) & = \alpha_{00} + \alpha_{10} \log c_0 \\
d^u_0 \mid c_0 &\sim\mbox{Bernoulli}(\pi_0^u)\\
\text{logit}(\pi^u_0) & = \gamma_{00} + \gamma_{10} \log c_0
\end{align*}
where $\nu^u_0$ and $\sigma^u_0$ are the mean and standard deviation for $u_0$ given $u_0\neq \bm 1$ and $c_0$, while $\pi_0^u$ is the probability of having a utility value of one given $c_0$. We use logistic transformations to define a linear dependence for $p(u_0 \mid c_0, u_0\neq1)$ and include the costs on the log scale to improve the fit of the~model.


At time $j=1,2$, we extend the approach illustrated for $j=0$, and make a first-order Markov~assumption. For the costs we have:
\begin{align*}
c_j \mid d^c_j=0, c_{j-1},u_{j-1} & \sim \text{LogNormal}\left(\nu_j^c,\tau^c_j\right)\\
\nu^c_j & = \beta_{0j} + \beta_{1j}\log c_{j-1} + \beta_{2j}u_{j-1}\\
d^c_j \mid c_{j-1},u_{j-1} &\sim\mbox{Bernoulli}(\pi_j^c)\\
\text{logit}(\pi^c_j) & = \zeta_{0j} + \zeta_{1j}\log c_{j-1}+ \zeta_{2j}u_{j-1}.
\end{align*}
Similarly to time $j=0$, the mean, standard deviation and probability parameters for the costs at time $j$ are indicated with $\nu^c_j,\tau^c_j$ and $\pi^c_j$. The regression parameters~$\bm \beta_j=(\beta_{0j},\beta_{1j},\beta_{2j})$ and $\bm \zeta_j=(\zeta_{0j},\zeta_{1j},\zeta_{2j})$ capture the dependence between costs at $j$ and the costs and utilities at $j-1$, for the non-zero and zero components, respectively. The model for the utilities is:
\begin{align*}
u_j \mid  d^u_j=0,  c_{j},u_{j-1} & \sim \text{Beta}\left(\nu_j^u,\sigma^u_j\right)\\
\text{logit}(\nu^u_j) & = \alpha_{0j} + \alpha_{1j}\log c_{j} + \alpha_{2j}u_{j-1}\\
d^u_j \mid c_{j},u_{j-1} &\sim\mbox{Bernoulli}(\pi_j^u)\\
\text{logit}(\pi^u_j) & = \gamma_{0j} + \gamma_{1j}\log c_{j}+ \gamma_{2j}u_{j-1}.
\end{align*}
We denote with $\nu^u_j,\sigma^u_j$ and $\pi^u_j$ the mean, standard deviation and probability parameters for the utilities at time $j$, and with $\bm \alpha_j=(\alpha_{0j},\alpha_{1j},\alpha_{2j})$ and $\bm \gamma_j=(\gamma_{0j},\gamma_{1j},\gamma_{2j})$ the regression parameters that capture the dependence between utilities at $j$ and costs at $j$ and utilities at $j-1$.

For all parameters in the model we specify vague prior distributions. Specifically a Normal with a large variance on the appropriate scale for the regression parameters and Uniform over a large positive range for the standard deviations. 
We implement the model and derive the marginal cost and utility means at each time $j$ through Monte Carlo Integration. First, we fit the model separately to the completers ($\bm r=\bm 1$) and the joint set of all other patterns ($\bm r \neq \bm 1$) for $t=1,2$. Second, at each iteration of the posterior distribution, we generate a large number of samples for $\bm y_{ij}=(c_{ij},u_{ij})$ based on the posterior values for the parameters of the utility and cost models in the MCMC output. Third, we approximate the posterior distribution of the marginal means for each $\bm r$ by taking the expectation over these sampled values at each iteration. Finally, we derive the overall marginal means~$\bm \mu_{jt}=(\mu^{c}_{jt},\mu^{u}_{jt})$ as weighted averages across the marginal means in each pattern, using the posterior $\bm \psi^{\bm r}_t$ as weights.
 
\section{Identifying Restrictions and Sensitivity Parameters}\label{SA}
Identifying restrictions provide a convenient approach to identify the extrapolation distribution and conduct sensitivity analysis. 
In short, identifying restrictions correspond to assumptions about $p(\bm y,\bm r)$, which link the observed data distribution $p(\bm y_{obs},\bm r)$ to the extrapolation distribution $p(\bm y_{mis} \mid \bm y_{obs},\bm r)$. It can be useful to specify a single identifying restriction as a benchmark assumption and consider interpretable deviations from that benchmark to assess how inferences are driven by our assumptions \citep{Linero2017}. As mentioned in Section~\ref{model}, we consider partial restrictions that do not fully identify the joint $p(\bm y,\bm r)$ but, in our setting, allow us to identify the posterior distribution of the marginal means. 


Sensitivity parameters $(\bm \Delta)$ are often embedded within identifying restrictions to assess the impact of alternative missingness assumptions on the quantities of interest. We choose $\bm \Delta_j=(\Delta^c_j,\Delta^u_j)$ to be time-specific location shifts at the marginal mean in each pattern \citep{Daniels2000}. Specifically, we identify the marginal mean of the missing responses in each pattern $\bm{y}^{\bm r}_{mis}$ by averaging across the corresponding components that are observed $\bm{y}^{{\bm r}^\prime}_{obs}({\mathcal{J}_{\bm r}^{{\bm r}^\prime}})$ for ${\bm r}^\prime\neq \bm 1$ and add the sensitivity parameters~$\bm \Delta_j$.
\begin{equation*}
\mbox{E}\left[\bm{y}^{\bm r}_{mis} \mid \bm r \right]=\mbox{E} \left[ \underset{{\bm r}^\prime \neq \bm 1, \mathcal{J}^{{\bm r}^\prime}_{\bm r}}{\mbox{E}}\left[ \bm{y}^{{\bm r}^\prime}_{obs}({\mathcal{J}_{\bm r}^{{\bm r}^\prime}})+\bm \Delta_j \mid {\bm r}^\prime \right] \right],
 \end{equation*}
for $j \in \{0,1,2\}$. 
As a reasonable benchmark assumption we set $\bm \Delta_j=\bm 0$. We then explore departures from this benchmark based on the assumption that subjects with a missing value at time $j$ are more likely to have a lower utility and a higher cost compared with those who are observed at the same time but were not a completer. We calibrate the priors on $\bm \Delta_j$ using the observed standard deviations for costs and utilities at each time $j$ to define the amplitude of the departures from~$\bm \Delta_j=\bm 0$.

\section{Application to the PBS Study}\label{results}

\subsection{Computation}
We fitted the model using \texttt{JAGS}, \citep{Plummer}, a software specifically designed for the analysis of Bayesian models using Markov Chain Monte Carlo (MCMC) simulation \citep{Brooks}, which can be interfaced with \texttt{R} through the package \texttt{R2jags} \citep{Su}. Samples from the posterior distribution of the parameters of interest generated by \texttt{JAGS} and saved to the \texttt{R} workspace are then used to produce summary statistics and plots. We ran two chains with 20,000 iterations per chain, using a burn-in of 5,000, for a total sample of 30,000 iterations for posterior inference. For each unknown quantity in the model, we assessed convergence and autocorrelation of the MCMC simulations using diagnostic measures including the \textit{potential scale reduction factor} and the \textit{effective sample size} \citep{Gelman2}. 

In the non-completers pattern ($\bm r \neq \bm 1$), we set to $0$ the regression parameters ($\zeta_{11}$,$\zeta_{21}$) and ($\gamma_{10}$,$\gamma_{11}$,$\gamma_{21}$) for the model fitted to the control and intervention group, respectively. This simplification was required because, among the non-completers, there is only one observed $c_j=0$  at time $j=1$ in the control group and one observed $u_j=1$ at time $j=\{0,1\}$  in the intervention group. We therefore drop from the model the dependence between the probabilities of having a structural value at these times and the variables at the previous or same times to ensure the convergence of the algorithm and avoid identifiability~problems.


\subsection{Model Assessment}\label{assessment}
We computed the Deviance Information Criterion \citep[DIC;][]{Spiegelhalter} to assess the fit of the model with respect to an alternative parametric specification, where the LogNormal distributions are replaced with Gamma distributions for the cost variables. The DIC is a measure of comparative predictive ability based on the model deviance and a penalty for model complexity known as effective number of parameters ($p_D$). When comparing a set of models based on the same data, the one associated with the lowest DIC is the best-performing, among those assessed. There are different ways of constructing the DIC in the presence of missing data, which means that its use and interpretation are not straightforward \citep{Celeux,Daniels2008,Masonb}. In our analysis, we consider a DIC based on the observed data under MAR as its value does not depend on the values of the sensitivity parameters \citep{Wang2011}. Because the sampling distribution of the observed data was not available in closed form, we computed it using Monte Carlo integration. Results between the two alternative specifications are reported in~Table~\ref{tt}.
\begin{center}
TABLE 2 HERE
\end{center}
The DIC components for the costs are systematically lower when LogNormal distributions are used compared with Gamma distributions (lower values shown in italics in Table~\ref{tt}), and result in an overall better fit to the data for the first model.

We also assess the absolute fit of the model using posterior predictive checks based on observed data replications \citep{Xu2016}. We sample from the posterior predictive distribution $p(\tilde{\bm y},\tilde{\bm r}\mid \bm y^{\bm r}_{obs},\bm r,\bm \omega)$. Conditional on the replicated patterns~$\tilde{\bm r}$, we define the replicated observed data in each pattern as~$\tilde{ \bm y}^{\tilde{\bm r}}_{obs}=\{\tilde{\bm y}_j : \tilde{\bm r}_j=\bm 1\}$, that is the components of $\tilde{\bm y}$ for which the corresponding missing data indicators at time~$j$ in the replicated patterns $\tilde{\bm r}$ are equal to one.

We compute the rank correlations between each pair of variables for each replicated dataset, and compare them with the corresponding values from the real dataset. The results, shown in Figure~\ref{corr}, suggest that the proposed parametric model captures most of the correlations well both in the control (panel a) and intervention (panel b) group.
\begin{center}
FIGURE 1 HERE
\end{center}

\subsection{Priors on Sensitivity Parameters}\label{priors}
We consider three alternative sets of priors on $\bm \Delta_j=(\Delta_j^u,\Delta_j^c)$, calibrated based on the variability in the observed data at each time $j$.
The three types of priors used are the following:
\begin{itemize}
\item \Large $\bm \Delta^{\text{flat}}$: \normalsize Flat between 0 and twice the observed standard deviation:
\begin{equation*}
\Delta^c_j \sim \text{Uniform}[0,2\; \text{sd}(c_j)] \;\;\; \text{and} \;\;\; \Delta^u_j \sim \text{Uniform}[-2 \; \text{sd}(u_j),0]
\end{equation*} 
\item \Large $\bm \Delta^{\text{skew0}}$: \normalsize Skewed towards values closer to 0, over the same range as $\bm \Delta^{\text{flat}}$:
\begin{equation*}
\Delta^c_j = 2\; \text{sd}(c_j) \left(1-\sqrt{\text{Uniform}[0,1]}\right) \;\;\; \text{and} \;\;\; \Delta^u_j = -2\text{\;sd}(u_j) \left(1-\sqrt{\text{Uniform}[0,1]}\right)\\
\end{equation*}
\item \Large $\bm \Delta^{\text{skew1}}$: \normalsize Skewed towards values far from 0, over the same range as $\bm \Delta^{\text{flat}}$:
\begin{equation*}
\Delta^c_j = 2\;\text{sd}(c_j) \left(\sqrt{\text{Uniform}[0,1]}\right) \;\;\; \text{and} \;\;\; \Delta^u_j =-2\text{\;sd}(u_j)\left(\sqrt{\text{Uniform}[0,1]}\right)\\
\end{equation*}
\end{itemize}
where $\text{sd}(u_j)$ and $\text{sd}(c_j)$ are the standard deviations computed on the observed utilities and costs at time $j$ for $\bm r \neq \bm 1$. We choose these priors because we believe that departures from $\bm \Delta_j=0$  for both outcomes are not likely to be larger than twice the observed standard~deviations at each time $j$. 

\subsection{Results}\label{res}
Figure~\ref{res_mean2} compares the posterior means and 95\% highest posterior density (HPD) credible intervals for $\bm \mu_{jt}=(\mu^u_{jt},\mu^c_{jt})$ obtained from fitting the model under six alternative scenarios: completers (CC), all cases assuming ignorability (MAR), and using the extrapolation factorisation under the benchmark ($\bm \Delta_j=\bm 0$) and three departure scenarios~($\bm \Delta^{\text{flat}}$,$\bm \Delta^{\text{skew0}}$,$\bm \Delta^{\text{skew1}}$). Since baseline costs are fully observed, only the estimates under CC and MAR are shown for $\bm \mu^c_0$. Results associated with the control and intervention group are indicated in red and blue,~respectively.
\begin{center}
FIGURE 2 HERE
\end{center}
The distributions of both $\bm \mu^u_j$ and $\bm \mu^c_j$ show values that are higher in the intervention compared with the control at each time $j$, and are similar across CC, MAR and $\bm \Delta=\bm0$. However, under the other nonignorable scenarios, mean utilities/costs are on average $3\%$ ($\bm \Delta^{\text{flat}}$), $4\%$ ($\bm \Delta^{\text{skew0}}$) and $5\%$ ($\bm \Delta^{\text{skew1}}$) lower/higher compared with $\bm \Delta=\bm 0$ in the control group. In the intervention group, mean utilities/costs are on average $1\%$ ($\bm \Delta^{\text{flat}}$), $1.5\%$ ($\bm \Delta^{\text{skew0}}$) and $2.5\%$ ($\bm \Delta^{\text{skew1}}$) lower/higher compared with~$\bm \Delta=\bm 0$.

We then derived the QALYs and total costs means $\mu_{et}$ and $\mu_{ct}$ by applying the formulae in Equation~\ref{QALYs} to the cost and utility marginal means $\bm \mu_{jt}$ obtained from the model. We also compare the estimates derived from our approach with those obtained from a cross-sectional model fitted on $e_i$ and $c_i$, computed only based on the completers in the study, as a standard approach used in trial-based analyses (CS). This model was specified following the approach used in the original analysis of the PBS study (assuming bivariate normality and including baseline adjustments), but implemented within a Bayesian framework.

Table~\ref{res_bar} shows the posterior means and 95\% HPD credible intervals associated with the targeted quantities under all~scenarios for both treatment groups.
\begin{center}
TABLE 3
\end{center}
Results under CS are systematically lower for both mean parameters with respect to those from any other scenario. Compared with CC, we respectively observe a decrease of 38\% and 21\% in the control and intervention group for $\mu_{et}$ and of 10\% and 5\% for $\mu_{ct}$. 

Across the other scenarios, variations of similar amplitude and with the same sign to those for $\bm \mu_{jt}$ affect the QALYs and total cost mean estimates. In the control, compared with $\bm \Delta=\bm 0$, mean QALYs and total costs show decrements between $2.8\%$ and $5.7\%$ and increments between $3.5\%$ and $7\%$ with respect to the three nonignorable scenarios, respectively. In the intervention, the corresponding decreases in mean QALYs are between $1\%$ and $2\%$ while the increases in mean total costs are between $1.2\%$ and~$2.3\%$ .

\section{Economic Evaluation}\label{evaluation}
We complete the analysis by assessing the cost-effectiveness of the new intervention with respect to the control, comparing the results under the cross-sectional (CS), complete case (CC), ignorable (MAR), benchmark nonignorable ($\bm \Delta=\bm 0$) and the three alternative nonignorable departure scenarios. We specifically rely on the examination of the Cost-Effectiveness Plane \citep[CEP;][]{Black} and the Cost-Effectiveness Acceptability Curve \citep[CEAC;][]{VanHout} to summarise the economic~analysis. 
\begin{center}
FIGURE 3 HERE
\end{center}
The CEP (Figure~\ref{CEAC}, panel a) is a graphical representation of the joint distribution for the population average effectiveness and costs increments between the two arms, indicated respectively as $\mu_{e2}-\mu_{e1}$ and $\mu_{c2}-\mu_{c1}$. We show the results only under three scenarios (light blue for CC, light green for MAR and light red for $\bm \Delta^{\text{flat}}$) for clarity and visualisation purposes. The results for the other nonignorable scenarios are available in the Web Appendix. The slope of the straight line crossing the plane is the ``willingness to pay'' threshold (often indicated as $k$). This can be considered as the amount of budget the decision-maker is willing to spend to increase the health outcome of one unit and effectively is used to trade clinical benefits for money. Points lying below this straight line fall in the so-called \textit{sustainability area} \citep{Baioa} and suggest that the active intervention is more cost-effective than the control. In the graph, we also show the Incremental Cost-Effectiveness Ratio (ICER) computed under each scenario, as darker coloured dots. This is defined as 
$$\mbox{ICER}=\frac{\mbox{E}[\mu_{c2}-\mu_{c1}]}{\mbox{E}[\mu_{e2}-\mu_{e1}]}$$ 
and quantifies the cost per incremental unit of effectiveness. For all three scenarios almost all samples fall in the North-East quadrant and are associated with positive ICERs. This suggests that the intervention is likely to produce both QALY gains and cost savings. The ICER under $\bm \Delta^{\text{flat}}$ falls in the sustainability area and indicates a more positive cost-effective assessment for the new intervention compared with CC and MAR.

The CEAC (Figure~\ref{CEAC}, panel b) is obtained by computing the proportion of points lying in the sustainability area upon varying the willingness to pay threshold $k$. Based on standard practice in routine analyses, we consider a range for $k$ up to \pounds{40,000} per QALY gained.  The CEAC estimates the probability of cost-effectiveness, thus providing a simple summary of the uncertainty associated with the ``optimal'' decision-making suggested by the ICER. The results under CC and MAR are reported using blue and green solid lines, respectively. In addition, the results derived under nonignorability are reported using different coloured dashed~lines. 

The CEACs under CC, MAR and the benchmark scenarios show a similar trend and indicate a probability of cost-effectiveness below $0.65$ of the new intervention for values of $k$ up to $\pounds{40,000}$. However, under the other scenarios, the curve is shifted upwards by an average probability of $0.2$ ($\bm \Delta^{\text{flat}}$), $0.15$ ($\bm \Delta^{\text{skew0}}$) and $0.25$ ($\bm \Delta^{\text{skew1}}$) and suggests a more favourable cost-effectiveness assessment. The CEAC plot shows that results are sensitive to the assumptions about the missing values, which can lead to a considerable change in the output of the decision process and the cost-effectiveness conclusions.
 
We finally compare the economic results under our longitudinal approach with respect to those derived from a typical cross-sectional model (CS). Figure~\ref{CEAC2} shows the CEPs (panel a) associated with the CS, CC and MAR scenarios, respectively indicated with red, blue and green coloured dots. In the CEACs (panel b), in addition to the probability values associated with these scenarios (solid lines), the results from $\bm \Delta^{\text{flat}}$ are indicated with a dashed line.
\begin{center}
FIGURE 4 HERE
\end{center} 
The distribution of the posterior samples in the CEP (Figure~\ref{CEAC2}, panel a) show some differences between the scenarios with the ICER; CS is the lowest among those compared. In the CEAC (Figure~\ref{CEAC2}, panel b), the acceptability curve for CS is higher than those for CC and MAR for most willingness to pay values but remains systematically lower with respect to the $\bm \Delta^{\text{flat}}$,  $\bm \Delta^{\text{skew0}}$ and  $\bm \Delta^{\text{skew1}}$ (in Figure~\ref{CEAC2} we only show the results for $\bm \Delta^{\text{flat}}$ for clarity).

\section{Discussion}\label{conclusions} 
Missingness represents a threat to economic evaluations as, when dealing with partially-observed data, any analysis makes assumptions about the missing values that cannot be verified from the data at hand. Trial-based analyses are typically conducted on cross-sectional quantities, e.g.~QALYs and total costs, which are derived based only on the observed data from the completers in the study. This is an inefficient and likely biased approach, unless the completers are a random sample of all study participants, because data from any partially-observed subject is lost. A further concern is that routine analyses typically rely on standard models that ignore or at best fail to properly account for potentially important features in the data such as correlation, skewness, and the presence of structural~values.

In this paper, we have proposed an alternative approach for conducting parametric Bayesian inference under nonignorable missingness for a longitudinal bivariate outcome in health economic evaluations, while accounting for typical data features such as skewness and presence of structural values in both utilities and costs. The analysis of the PBS data shows the benefits of using our approach compared with a standard cross-sectional model and a considerable impact of alternative MNAR assumptions on the final decision-making conclusions, suggesting a more cost-effective intervention compared with the results obtained under ignorability~(MAR). 

We relied on the extrapolation factorisation, within a pattern mixture approach, and handled the sparsity of the data in most patterns by collapsing the non-completers together when fitting the model. We identified the extrapolation distribution only up to the marginal mean with partial identifying restrictions using the marginal means estimated from the incompleters. As an alternative approach, we could have used the marginal mean estimates from the completers, but we considered those of the incompleters as a more reasonable default MNAR assumption. Next, we used sensitivity parameters to characterise the uncertainty about the missing data within each pattern. Alternative prior choices, calibrated in different ways using the observed data, were chosen for the sensitivity parameters and the robustness of the results across these scenarios was assessed.

An area for future work is to increase the flexibility of our approach through a semi-parametric specification for the observed data distribution, which would allow a weakening of the model assumptions and likely further improve the fit of the model to the observed data. As for the extrapolation distribution, alternative identifying restrictions that introduce the sensitivity parameters via the conditional mean (rather than the marginal mean) could be considered, and their impact on the conclusions assessed in sensitivity analysis.

\ack{Dr Michael J. Daniels was partially supported by the US NIH grant CA-183854.\\
Dr Gianluca Baio is partially supported as the recipient of an unrestricted research grant sponsored by Mapi Group at University College London.\\
Mr Andrea Gabrio is partially funded in his PhD programme at University College London by a research grant sponsored by The Foundation BLANCEFLOR Boncompagni Ludovisi, n\'{e}e Bildt.

\bibliographystyle{rss}
\bibliography{long_model}

\begin{thebibliography}{44}
\expandafter\ifx\csname natexlab\endcsname\relax\def\natexlab#1{#1}\fi
\expandafter\ifx\csname url\endcsname\relax
  \def\url#1{\texttt{#1}}\fi
\expandafter\ifx\csname urlprefix\endcsname\relax\def\urlprefix{URL: }\fi

\bibitem[{Baio(2012)}]{Baioa}
Baio, G. (2012) \textit{Bayesian Methods in Health Economics}.
\newblock University College London, London, UK: Chapman and Hall/CRC.

\bibitem[{Baio\vspace{0mm}(2014)}]{Baio2014}
Baio\vspace{0mm}, G. (2014) Bayesian models for cost-effectiveness analysis in
  the presence of structural zero costs.
\newblock \textit{Statistics in Medicine}, \textbf{33}, 1900--1913.

\bibitem[{Black(1990)}]{Black}
Black, W. (1990) A graphic representation of cost-effectiveness.
\newblock \textit{Medical Decision Making}, \textbf{10}, 212--214.

\bibitem[{Briggs(2000)}]{Briggs2000}
Briggs, A. (2000) Handling uncertainty in cost-effectiveness models.
\newblock \textit{PharmacoEconomics}, \textbf{22}, 479--500.

\bibitem[{Briggs et~al.(2006)Briggs, Schulpher and Claxton}]{Briggs2006}
Briggs, A., Schulpher, M. and Claxton, K. (2006) \textit{Decision Modelling for
  Health Economic Evaluation}.
\newblock Oxford, UK: Oxford university press.

\bibitem[{Brooks et~al.(2011)Brooks, Gelman, Jones and Meng}]{Brooks}
Brooks, S., Gelman, A., Jones, G. and Meng, X. (2011) \textit{Handbook of
  {Markov Chain Monte Carlo}}.
\newblock CRC press.

\bibitem[{Celeux et~al.(2006)Celeux, Forbes, Robert and Titterington}]{Celeux}
Celeux, G., Forbes, S., Robert, C. and Titterington, D. (2006) Deviance
  information criteria for missing data models.
\newblock \textit{Bayesian Analysis}, \textbf{1}, 651--674.

\bibitem[{Claxton(1999)}]{Claxtonb}
Claxton, K. (1999) The irrelevance of inference: a decision making approach to
  stochastic evaluation of health care technologies.
\newblock \textit{Journal of Health Economics}, \textbf{18}, 342--364.

\bibitem[{Cooper et~al.(2003)Cooper, Sutton, Mugford and Abrams}]{Cooper}
Cooper, N., Sutton, A., Mugford, M. and Abrams, K. (2003) Use of {Bayesian
  Markov Chain Monte Carlo} methods to model cost-of-illness based on general
  recommended guidelines.
\newblock \textit{Medical Decision Making}, \textbf{23}, 38--53.

\bibitem[{Daniels and Hogan(2000)}]{Daniels2000}
Daniels, M. and Hogan, J. (2000) Reparameterizing the pattern mixture model for
  sensitivity analysis under informative dropout.
\newblock \textit{Biometrics}, \textbf{56}, 1241--1248.

\bibitem[{Daniels\vspace{0mm} and Hogan(2008)}]{Daniels2008}
Daniels\vspace{0mm}, M. and Hogan, J. (2008) \textit{Missing Data in
  Longitudinal Studies: Strategies for Bayesian Modeling and Sensitivity
  Analysis}.
\newblock New York, US: Chapman and Hall.

\bibitem[{{European Medicines Agency}(2013)}]{Agency}
{European Medicines Agency} (2013) {Committee for Medicinal Products for Human
  Use (CHMP). Guideline on adjustment for baseline covariates}.
\newblock
  \url{http://www.ema.europa.eu/docs/en_GB/document_library/Scientific_guideline/2013/06/WC500144946.pdf}.

\bibitem[{Gabrio et~al.(2017)Gabrio, Mason and Baio}]{Gabrio}
Gabrio, A., Mason, A. and Baio, G. (2017) Handling missing data in within-trial
  cost-effectiveness analysis: A review with future recommendations.
\newblock \textit{PharmacoEconomics-Open}, \textbf{1}, 79--97.

\bibitem[{Gabrio et~al.(2018)Gabrio, Mason and Baio}]{Gabriob}
Gabrio, A., Mason, J. and Baio, G. (2018) A full {Bayesian} model to handle
  structural ones and missingness in economic evaluations from individual-level
  data.
\newblock \url{https://arxiv.org/abs/1801.09541}.

\bibitem[{Gaskins et~al.(2016)Gaskins, Daniels and Marcus}]{Gaskins}
Gaskins, J., Daniels, M. and Marcus, B. (2016) Bayesian methods for
  nonignorable dropout in joint models in smoking cessation studies.
\newblock \textit{Journal of the American Statistical Association},
  \textbf{111}, 1454--1465.

\bibitem[{Gelman et~al.(2004)Gelman, Carlin, Stern and Rubin}]{Gelman2}
Gelman, A., Carlin, J., Stern, H. and Rubin, D. (2004) \textit{{Bayesian Data
  Analysis - 2nd edition}}.
\newblock New York, NY: Chapman and Hall.

\bibitem[{Hassiotis et~al.(2018)Hassiotis, Poppe, Strydom, Vickerstaff, Hall,
  Crabtree, Omar, King, Hunter, Bosco, Biswas, Ratti, Blickwedel, Cooper, Howie
  and Crawford}]{Hassiotis2018}
Hassiotis, A., Poppe, M., Strydom, A., Vickerstaff, V., Hall, I., Crabtree, J.,
  Omar, R., King, M., Hunter, R., Bosco, A., Biswas, A., Ratti, V., Blickwedel,
  J., Cooper, V., Howie, W. and Crawford, M. (2018) Positive behaviour support
  training for staff for treating challenging behaviour in people with
  intellectual disabilities: a cluster rct.
\newblock \textit{Health Technology Assessment}, \textbf{22}.

\bibitem[{Jackson et~al.(2009)Jackson, Thompson and Sharples}]{Jackson}
Jackson, C., Thompson, S. and Sharples, L. (2009) Accounting for uncertainty in
  health economic decision models by using model averaging.
\newblock \textit{Journal of the Royal Statistical Society: Series A},
  \textbf{172}, 383--404.

\bibitem[{Leurent et~al.(2018)Leurent, Gomes and Carpenter}]{Leurent2018}
Leurent, B., Gomes, M. and Carpenter, J. (2018) Missing data in trial-based
  cost-effectiveness analysis: An incomplete journey.
\newblock \textit{Health Economics}.

\bibitem[{Linero and Daniels(2015)}]{Linero2015}
Linero, A. and Daniels, M. (2015) A flexible {Bayesian} approach to monotone
  missing data in longitudinal studies with nonignorable missingness with
  application to an acute schizophrenia clinical trial.
\newblock \textit{Journal of the American Statistical Association},
  \textbf{110}, 45--55.

\bibitem[{Linero\vspace{0mm} and Daniels(2018)}]{Linero2017}
Linero\vspace{0mm}, A. and Daniels, M. (2018) Bayesian approaches for missing
  not at random outcome data: The role of identifying restrictions.
\newblock \textit{Statistical Science}, \textbf{33}, 198--213.

\bibitem[{Little\vspace{0mm}(1994)}]{Little1994}
Little\vspace{0mm}, R. (1994) A class of pattern-mixture models for normal
  incomplete data.
\newblock \textit{Biometrika}, \textbf{81}, 471--483.

\bibitem[{Little\vspace{0mm} and Rubin(2002)}]{Little2002}
Little\vspace{0mm}, R. and Rubin, D. (2002) \textit{Statistical Analysis with
  Missing Data, Second Edition}.
\newblock New York: John Wiley and Sons.

\bibitem[{Manca et~al.(2005)Manca, Hawkins and Sculpher}]{Manca}
Manca, A., Hawkins, N. and Sculpher, M. (2005) Estimating mean {QALYs} in
  trial-based cost-effectiveness analysis: the importance of controlling for
  baseline utility.
\newblock \textit{Health Economics}, \textbf{14}, 487--496.

\bibitem[{Mason et~al.(2012)Mason, Richardson and Best}]{Masonb}
Mason, A., Richardson, S. and Best, N. (2012) Two-pronged strategy for using
  {DIC} to compare selection models with non-ignorable missing responses.
\newblock \textit{Bayesian Analysis}, \textbf{7}, 109--146.

\bibitem[{Molenberghs et~al.(1997)Molenberghs, Kenward and
  Lesaffre}]{Molenberghs1997}
Molenberghs, G., Kenward, M. and Lesaffre, E. (1997) The analysis of
  longitudinal ordinal data with non-random drop-out.
\newblock \textit{Biometrika}, \textbf{84}, 33--44.

\bibitem[{NICE(2013)}]{NICE2013}
NICE (2013) \textit{Guide to the Methods of Technological Appraisal}.
\newblock London, UK: NICE.

\bibitem[{Nixon and Thompson(2005)}]{Nixon}
Nixon, R. and Thompson, S. (2005) Methods for incorporating covariate
  adjustment, subgroup analysis and between-centre differences into
  cost-effectiveness evaluations.
\newblock \textit{Health Economics}, \textbf{14}, 1217--1229.

\bibitem[{Noble et~al.(2012)Noble, Hollingworth and Tilling}]{Noble2012}
Noble, S., Hollingworth, W. and Tilling, K. (2012) Missing data in trial-based
  cost-effectiveness analysis: the current state of play.
\newblock \textit{Health Economics}, \textbf{21}, 187--200.

\bibitem[{OHagan et~al.(2004)OHagan, McCabe, Hakehurst, Brennan, Briggs,
  Claxton, Fenwick, Fryback, Schulpher, Spiegelhalter and Willan}]{OHagan2004}
OHagan, A., McCabe, C., Hakehurst, R., Brennan, A., Briggs, A., Claxton, K.,
  Fenwick, E., Fryback, D., Schulpher, M., Spiegelhalter, D. and Willan, A.
  (2004) Incorporation of uncertainty in health economic modelling studies.
\newblock \textit{PharmacoEconomics}, \textbf{23}, 529--536.

\bibitem[{O'Hagan and Stevens(2001)}]{OHagan}
O'Hagan, A. and Stevens, J. (2001) A framework for cost-effectiveness analysis
  from clinical trial data.
\newblock \textit{Health Economics}, \textbf{10}, 303--315.

\bibitem[{Plummer(2010)}]{Plummer}
Plummer, M. (2010) {JAGS: Just Another Gibbs Sampler}.
\newblock \url{http://www-fis.iarc.fr/~martyn/software/jags/}.

\bibitem[{Rubin(1987)}]{Rubina}
Rubin, D. (1987) \textit{Multiple Imputation for Nonresponse in Surveys}.
\newblock New York, US: John Wiley and Sons.

\bibitem[{Scharfstein et~al.(1999)Scharfstein, Rotnitzky and
  Robins}]{Scharfstein1999}
Scharfstein, D., Rotnitzky, A. and Robins, J. (1999) Adjusting for nonignorable
  drop-out using semiparametric nonresponse models.
\newblock \textit{Journal of the American Statistical Association},
  \textbf{94}, 1135--1146.

\bibitem[{Sculpher et~al.(2005)Sculpher, Claxton, Drummond and
  McCabe}]{Sculpher}
Sculpher, M., Claxton, K., Drummond, M. and McCabe, C. (2005) Whither
  trial-based economic evaluation for health decision making?
\newblock \textit{Health Economics}, \textbf{15}, 677--687.

\bibitem[{Spiegelhalter et~al.(2004)Spiegelhalter, Abrams and
  Myles}]{Spiegelhalterb}
Spiegelhalter, D., Abrams, K. and Myles, J. (2004) \textit{Bayesian approaches
  to clinical trials and health-care evaluation}.
\newblock John Wiley and Sons.

\bibitem[{Spiegelhalter et~al.(2002)Spiegelhalter, Best, Carlin and van~der
  Linde}]{Spiegelhalter}
Spiegelhalter, D., Best, N., Carlin, B. and van~der Linde, A. (2002) {Bayesian}
  measures of model complexity and fit.
\newblock \textit{Journal of the Royal Statistical Society}, \textbf{64},
  583--639.

\bibitem[{Su and Yajima(2015)}]{Su}
Su, Y. and Yajima, M. (2015) {Package ‘R2jags’}.
\newblock \url{https://cran.r-project.org/web/packages/R2jags/index.html}.

\bibitem[{Thompson and Nixon(2005)}]{Thompson}
Thompson, S. and Nixon, R. (2005) How sensitive are cost-effectiveness analyses
  to choice of parametric distributions?
\newblock \textit{Medical Decision Making}, \textbf{4}, 416--423.

\bibitem[{Van~Asselt et~al.(2009)Van~Asselt, van Mastrigt, Dirksen, Arntz,
  Severens and Kessels}]{Asselt}
Van~Asselt, A., van Mastrigt, G., Dirksen, C., Arntz, A., Severens, J. and
  Kessels, A. (2009) How to deal with cost differences at baseline.
\newblock \textit{PharmacoEconomics}, \textbf{27}, 519--528.

\bibitem[{Van~Hout et~al.(1994)Van~Hout, Al, Gordon, Rutten and
  Kuntz}]{VanHout}
Van~Hout, B., Al, M., Gordon, G., Rutten, F. and Kuntz, K. (1994) Costs,
  effects and {C/E-Ratios} alongside a clinical trial.
\newblock \textit{Health Economics}, \textbf{3}, 309--319.

\bibitem[{Vansteelandt et~al.(2006)Vansteelandt, Goetghebeur, Kenward and
  Molenberghs}]{Vansteelandt2006}
Vansteelandt, S., Goetghebeur, E., Kenward, M. and Molenberghs, G. (2006)
  Ignorance and uncertainty regions as inferential tools in a sensitivity
  analysis.
\newblock \textit{Statistica Sinica}, \textbf{16}, 953--979.

\bibitem[{Wang and Daniels(2011)}]{Wang2011}
Wang, C. and Daniels, M. (2011) A note on {MAR}, identifying restrictions,
  model comparison, and sensitivity analysis in pattern mixture models with and
  without covariates for incomplete data.
\newblock \textit{Biometrics}, \textbf{67}, 810--818.

\bibitem[{Xu et~al.(2016)Xu, Chatterjee and Daniels}]{Xu2016}
Xu, D., Chatterjee, A. and Daniels, M. (2016) A note on posterior predictive
  checks to assess model fit for incomplete data.
\newblock \textit{Statistics in medicine}, \textbf{35}, 5029--5039.

\end{thebibliography}

\clearpage

\begin{table}
\caption{\label{patterns}Missingness patterns for the outcome $\bm y_j=(u_j,c_j)$ in the PBS study. For each pattern and treatment group, the number of subjects ($n_{\bm r t}$) and the observed mean responses at each time $j=0,1,2$ are reported. We denote the absence of response values or individuals within each pattern with --.}
\centering
\scalebox{0.85}{
\begin{tabular}{|c|cccccc|c|cccccc|c|}
\toprule
 \multicolumn{1}{|c}{}  & \multicolumn{6}{c|}{control ($t=1$)} & & \multicolumn{6}{c|}{intervention ($t=2$)}&\\
\multicolumn{1}{|c}{}   & $u_0$ & $c_0$ & $u_1$ & $c_1$ & $u_2$ & $c_2$ & $n_{\bm r 1}$ & $u_0$ & $c_0$ & $u_1$ & $c_1$ & $u_2$ & $c_2$ & $n_{\bm r 2}$\\
\midrule
$\bm r =\bm 1 $  & 1 & 1 & 1 & 1 & 1 & 1 & \multirow{2}{*}{108} & 1 & 1 & 1 & 1 & 1& 1& \multirow{2}{*}{96}\\ 
mean   & 0.678 & 1546 & 0.684 & 1527 & 0.680 & 1520 &  & 0.726 & 2818 & 0.771 & 2833 & 0.759 & 2878 & \\ 
$\bm r$  & 0 & 1 & 1 & 1 & 1 & 1 & \multirow{2}{*}{7} & 0 & 1 & 1 & 1 & 1& 1& \multirow{2}{*}{5}\\ 
mean   & -- & 1310 & 0.704 & 1440 & 0.644 & 1858 &  & -- & 2573 & 0.780 & 2939 & 0.849 & 2113 & \\ 
$\bm r$  & 1 & 1 & 0 & 1 & 1 & 1 & \multirow{2}{*}{4} & 1 & 1 & 0 & 1 & 1& 1& \multirow{2}{*}{1}\\ 
mean   & 0.709 & 1620 & -- & 1087 & 0.737 & 851 &  & 0.467 & 9649 & -- & 4828 & 0.259 & 4930 & \\ 
$\bm r$   & 1 & 1 & 1 & 1 & 0 & 1 & \multirow{2}{*}{2} & 1 & 1 & 1 & 1 & 0 & 1& \multirow{2}{*}{1}\\ 
mean   & 0.564 & 640 & 0.648 & 512 & -- & 286 &  & 0.817 & 3788 & 0.884 & 0 & -- & 0 & \\ 
$\bm r$   & 1 & 1 & 0 & 0 & 1 & 1 & \multirow{2}{*}{4} & 1 & 1 & 0 & 0 & 1 & 1 & \multirow{2}{*}{1}\\ 
mean   & 0.716 & 2834 & -- & -- & 0.634 & 679 &  & 0.501 & 3608 & -- & -- & 0.872 & 4781 & \\ 
$\bm r$  & 1 & 1 & 0 & 0 & 0 & 0 & \multirow{2}{*}{4} & 1 & 1 & 0 & 0 & 0 & 0 & \multirow{2}{*}{4}\\ 
mean   & 0.434 & 1528 & -- & -- & -- & -- &  & 0.760 & 3086 & -- & -- & -- & -- & \\ 
$\bm r$   & 0 & 1 & 0 & 1 & 1 & 1 & \multirow{2}{*}{2} & 0 & 1 & 0 & 1 & 1& 1& \multirow{2}{*}{0}\\ 
mean   & -- & 595 & -- & 397 & 0.483 & 69 &  & -- & -- & -- & -- & -- & -- & \\ 
$\bm r$   & 1 & 1 & 1 & 1 & 0 & 0 & \multirow{2}{*}{2} & 1 & 1 & 1 & 1 & 0 & 0 & \multirow{2}{*}{0}\\ 
mean  & 0.743 & 1434 & 0.705 & 1606 & -- & -- &  & -- & -- & -- & -- & -- & -- & \\ 
$\bm r$  & 1 & 1 & 0 & 1 & 0 & 1 & \multirow{2}{*}{3} & 1 & 1 & 0 & 1 & 0 & 1 & \multirow{2}{*}{0}\\ 
mean  & 0.726 & 1510 & -- & 432 & -- & 976 &  & -- & -- & -- & -- & -- & -- & \\ 
\bottomrule
\end{tabular}
}
\end{table}

\begin{table}
\caption{\label{tt}DIC and $p_D$ based on the observed data likelihood for each variable in the model. Two models are assessed either assuming LogNormal or Gamma distributions for the cost variables (lower DIC values shown in italics). Total DIC and $p_D$ are also reported at the bottom of the table.}
\centering
\scalebox{1}{
\begin{tabular}{c|cc|cc}
\toprule
& \multicolumn{2}{c|}{Gamma} &\multicolumn{2}{c}{LogNormal}\\
\midrule
variable & DIC & $p_D$ & DIC & $p_D$ \\ 
\midrule
$c_0$ & 2147.91  & 2.05 & \textit{2133.39} & 1.97\\ 
    $u_0 \mid c_0$ & -377.52  & 2.87  & -377.62 & 2.82\\ 
    $c_1 \mid c_0,u_0$ & 1904.53  & 4.16 & \textit{1827.45} & 4.13 \\ 
      $u_1 \mid u_0,c_1$ & -468.02  & 5.37 & -468.19  & 5.32 \\ 
  $c_2\mid c_1,u_1$ & 1913.69  & 4.65 & \textit{1856.23} & 4.36 \\ 
  $u_2 \mid u_1,c_2$ & -454.07  & 5.87 & -453.47  & 5.99 \\ 
  \midrule
 Total & 4667 & 25  & \textit{4518} & 25\\
\bottomrule
\end{tabular}
}
\end{table}

\begin{figure}[!h]
\centering
\subfloat[control]{\includegraphics[scale=0.55]{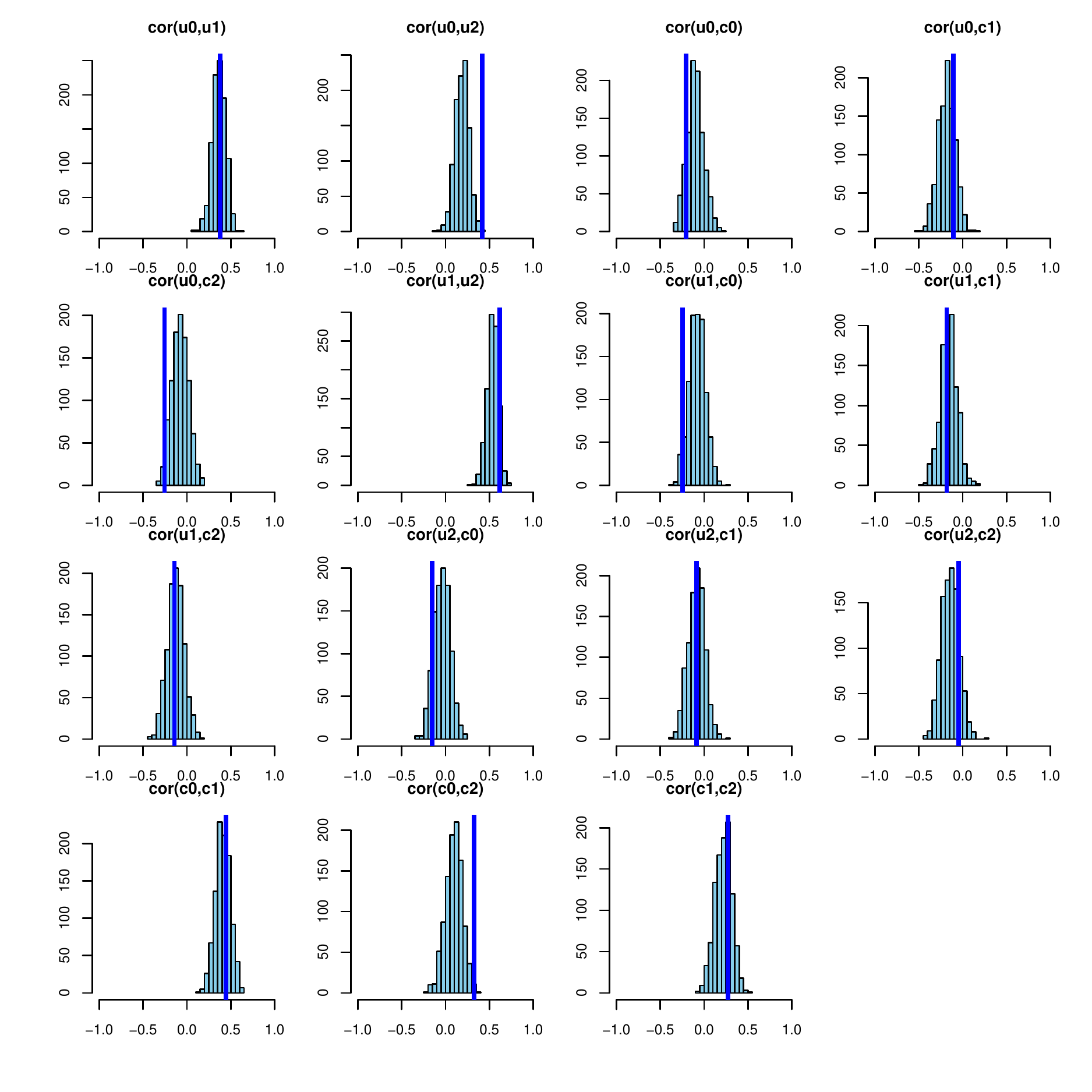}}\\
\subfloat[intervention]{\includegraphics[scale=0.55]{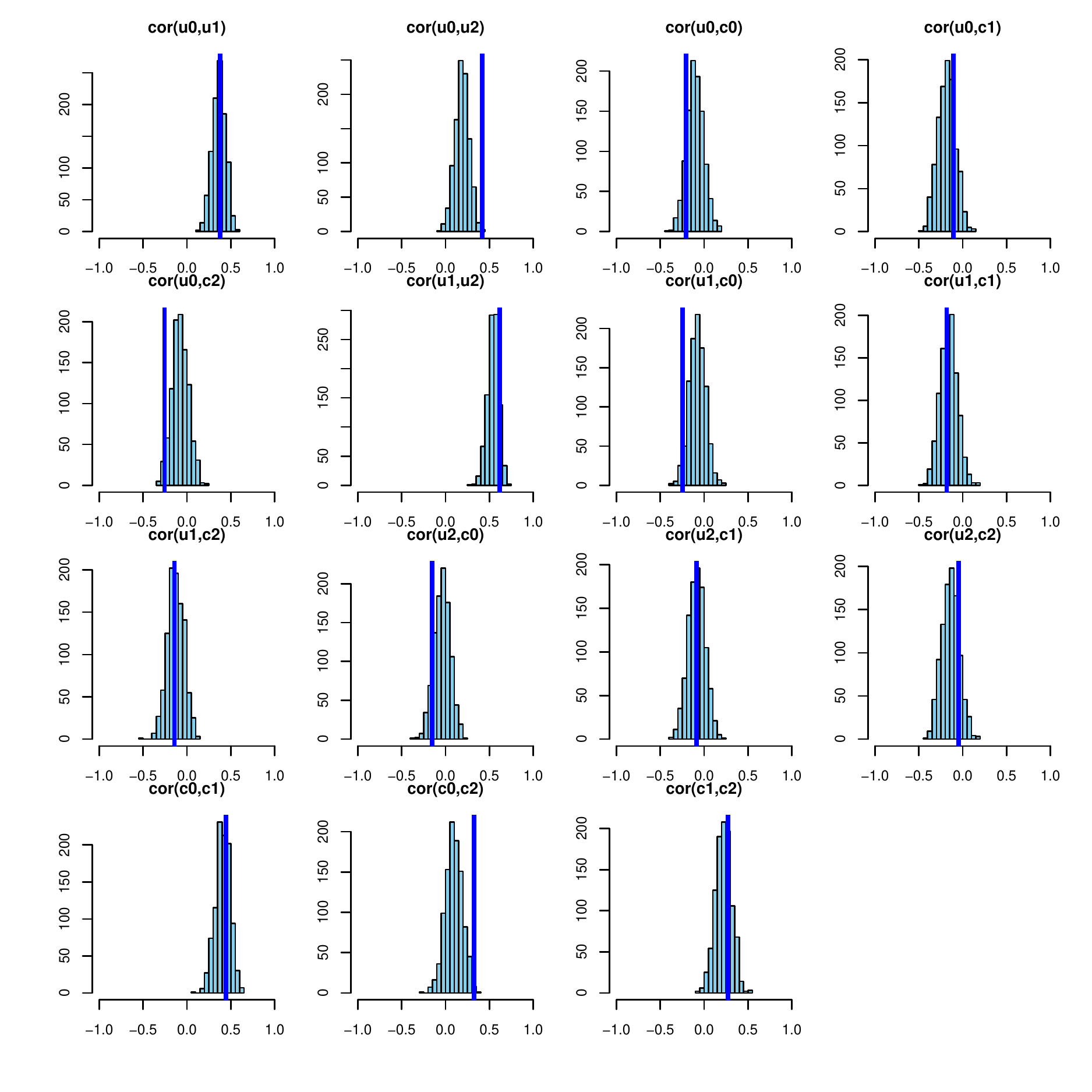}}
\caption{Posterior predictive distributions for the pairwise correlation between utilities and costs variables in the control (panel a) and intervention (panel b) arm across 1000 observed replicated datasets (light blue bars) compared with the estimates based on the observed data in the real dataset (vertical blue lines).}\label{corr}
\end{figure}

\clearpage

\begin{figure}[!h]
\centering
\subfloat{\includegraphics[scale=0.6]{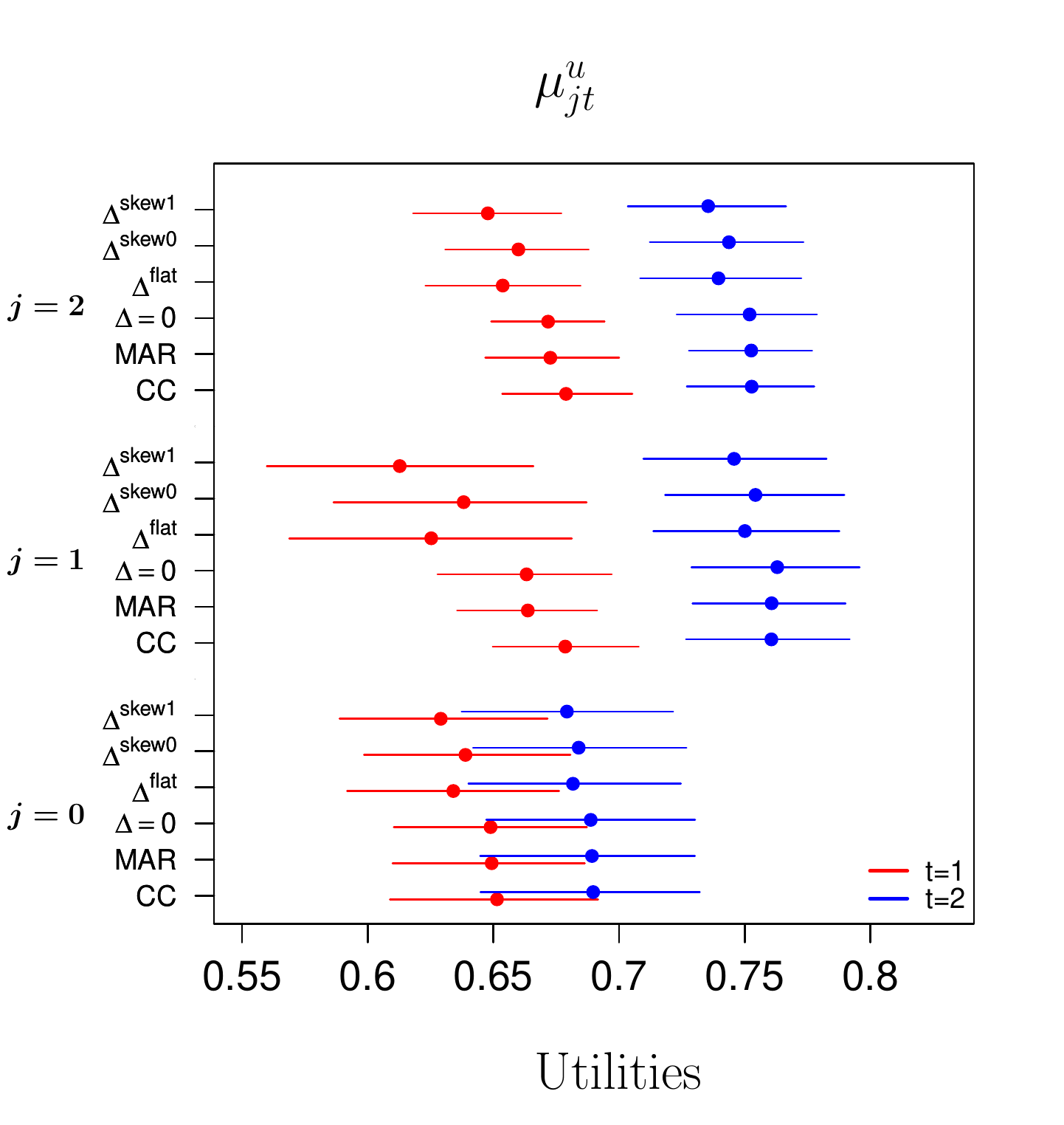}}
\subfloat{\includegraphics[scale=0.6]{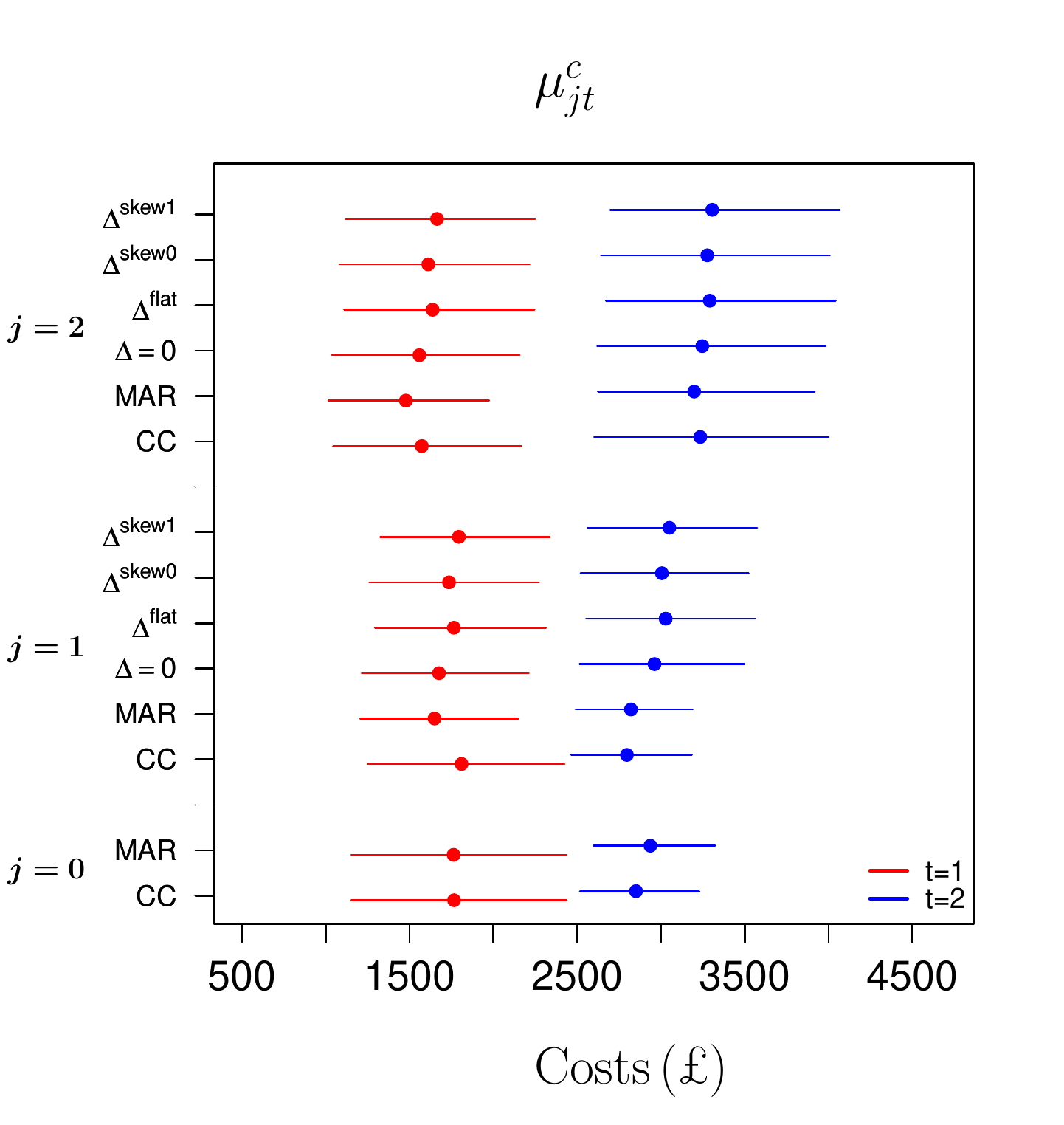}}
\caption{Posterior means and 95\% HPD intervals for the marginal utility and cost means in the control (red dots and lines) and intervention (blue dots and lines) group at each time $j$ in the study across alternative assumptions. Six scenarios are compared: completers (CC), ignorability (MAR), and nonignorability using the extrapolation factorisation under the benchmark assumption ($\bm\Delta=\bm0$) and under the three scenarios described in Section~\ref{priors} ($\bm\Delta^{\text{flat}}$,$\bm\Delta^{\text{skew0}}$,$\bm\Delta^{\text{skew1}}$). Since the baseline costs are fully observed in both groups, only the results under CC and MAR are displayed for $\bm \mu^c_0$.}\label{res_mean2}
\end{figure}

\clearpage

\begin{table}
\caption{\label{res_bar}Posterior means and 95\% HPD credible intervals for $\mu_{et}$ and $\mu_{ct}$ in the control ($t=1$) and intervention ($t=2$) group under alternative scenarios: Cross-Sectional (CS), CC, MAR,~$\bm \Delta=\bm 0$, $\bm \Delta^{\text{flat}}$,$\bm \Delta^{\text{skew0}}$ and $\bm \Delta^{\text{skew1}}$}
\centering
\scalebox{0.85}{
\begin{tabular}{c|cc|cc|cc|cc}
  \toprule
 \multirow{2}{*}{Scenario}   & \multicolumn{2}{c|}{$\mu_{e1}$} &  \multicolumn{2}{c|}{$\mu_{e2}$} &  \multicolumn{2}{c|}{$\mu_{c1}$} &  \multicolumn{2}{c}{$\mu_{c2}$} \\   \cmidrule{2-9}
  & mean & 95\% CI & mean & 95\% CI & mean & 95\% CI & mean & 95\% CI \\ 
  \midrule
  CS & 0.487 & (0.452; 0.524) & 0.611 & (0.570; 0.651) & 3073 & (2188; 3915) & 5768 & (5115; 6413) \\[0.25em] 
CC & 0.672 & (0.653; 0.691) & 0.741 & (0.720; 0.762) & 3382 & (2583; 4246) & 6031 & (5281; 6889) \\[0.25em] 
  MAR & 0.662 & (0.645; 0.681) & 0.741 & (0.721; 0.760) & 3125 & (2483; 3846) & 6018 & (5314; 6806) \\[0.25em]
  $\bm \Delta = \bm 0$ & 0.662 & (0.641; 0.682) & 0.742 & (0.721; 0.763) & 3233 & (2489; 4041) & 6208 & (5364; 7142) \\[0.25em]
 $\bm \Delta^{\text{flat}}$ & 0.635 & (0.601; 0.666) & 0.730 & (0.707; 0.753) & 3400 & (2616; 4196) & 6318 & (5462; 7271) \\[0.25em]
 $\bm \Delta^{\text{skew0}}$ & 0.644 & (0.615; 0.672) & 0.734 & (0.712; 0.756) & 3345 & (2605; 4173) & 6281 & (5409; 7200) \\[0.25em]
$\bm \Delta^{\text{skew1}}$ & 0.626 & (0.594; 0.656) & 0.727 & (0.703; 0.749) & 3457 & (2678; 4250) & 6355 & (5522; 7332) \\ 
   \bottomrule
\end{tabular}
}
\end{table}

\clearpage

\begin{figure}[!h]
\centering
\subfloat[]{\includegraphics[scale=0.45]{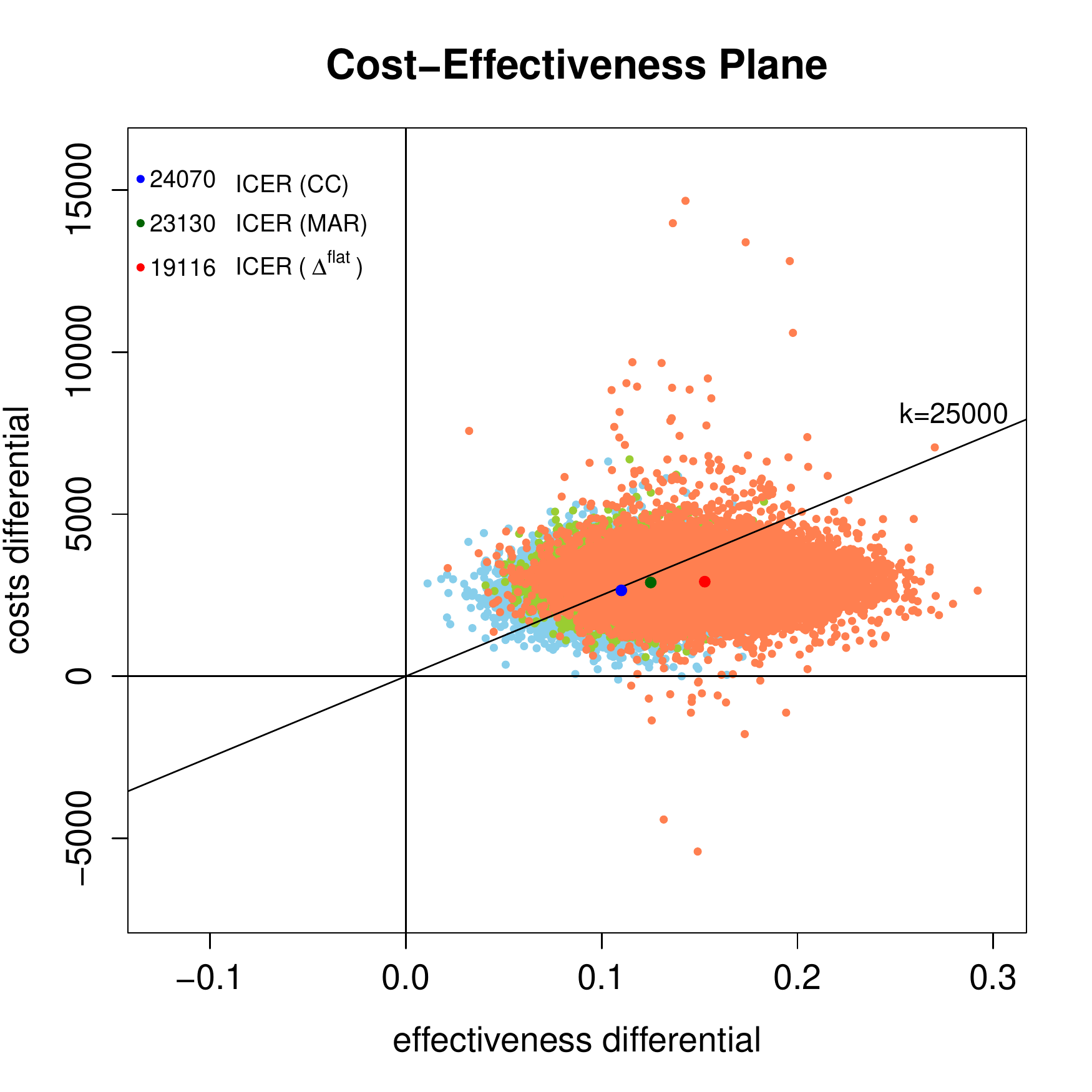}}
\subfloat[]{\includegraphics[scale=0.45]{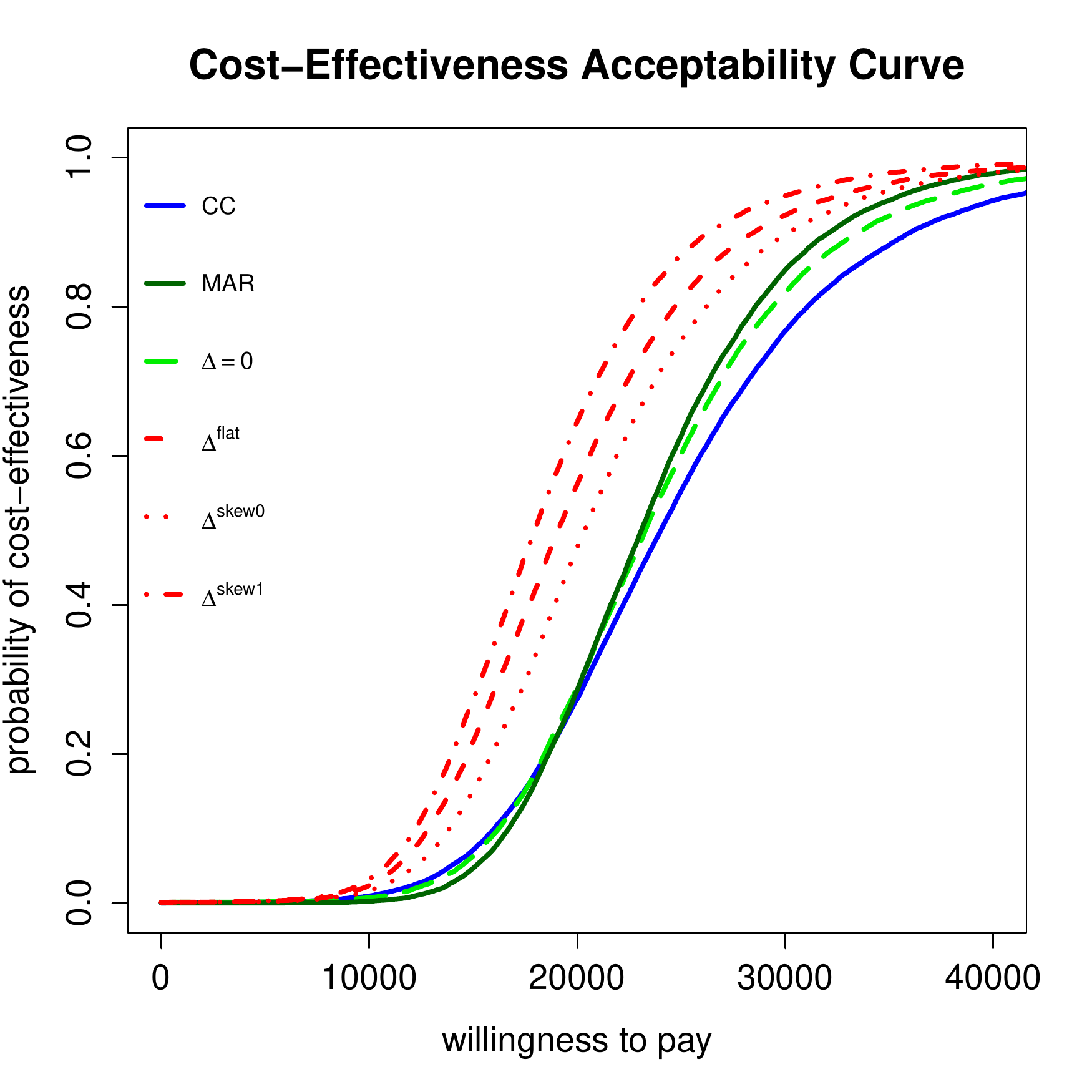}}
\caption{CEPs (panel a) and CEACs (panel b) associated with alternative missingness scenarios. In the CEPs, the ICERs based on the results from the complete cases (CC), MAR and $\bm \Delta^{\text{flat}}$ are indicated with corresponding darker coloured dots, while the portion of the plane on the right-hand side of the straight line passing through the plot (evaluated at $k=\text{\pounds{}}25,000$) denotes the sustainability area. For the CEACs, in addition to the results under CC and MAR (solid lines), the probability values for the alternative scenarios are represented with different coloured dashed lines.}\label{CEAC}
\end{figure}

\clearpage

\begin{figure}[!h]
\centering
\subfloat[]{\includegraphics[scale=0.45]{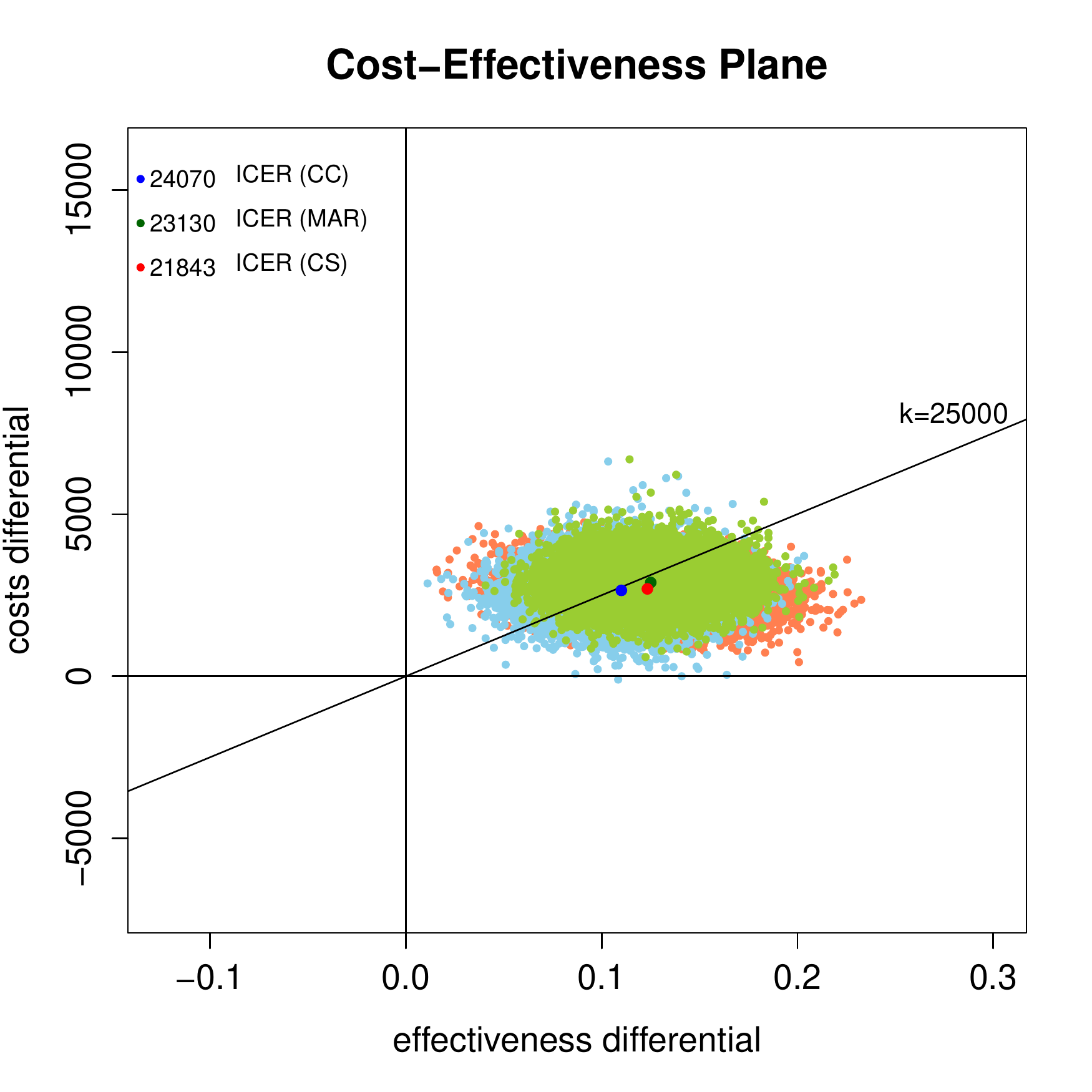}}
\subfloat[]{\includegraphics[scale=0.45]{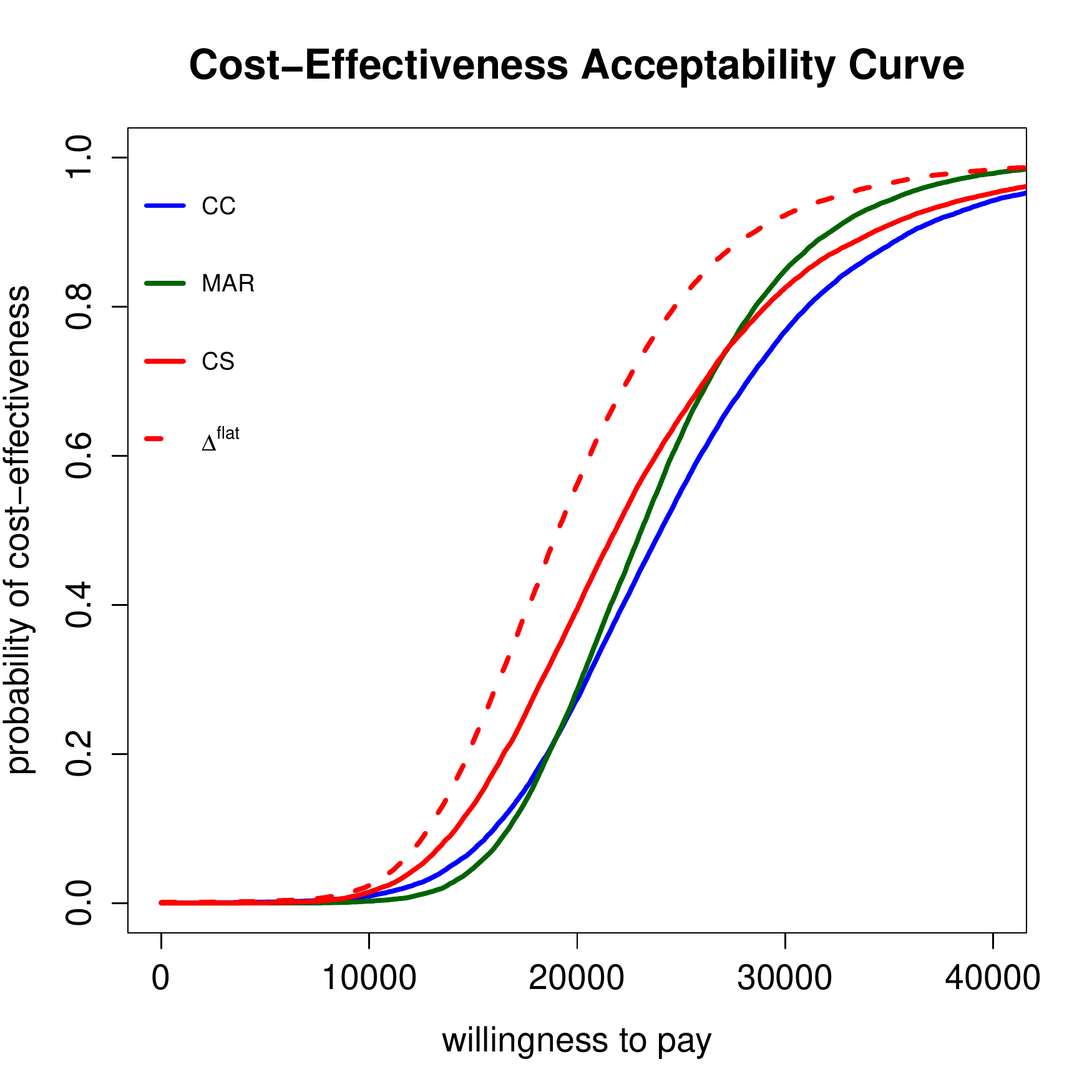}}
\caption{CEPs (panel a) and CEACs (panel b) associated with alternative scenarios. In the CEPs, the ICERs based on the results from the cross-sectional model (CS), complete cases (CC) and MAR are indicated with corresponding darker coloured dots, while the portion of the plane on the right-hand side of the straight line passing through the plot (evaluated at $k=\text{\pounds{}}25,000$) denotes the sustainability area. For the CEACs, in addition to the results under CS, CC and MAR (solid lines), the probability values for $\bm \Delta^{\text{flat}}$ are represented with a dashed line.}\label{CEAC2}
\end{figure}




\end{document}